\documentclass[sn-mathphys,Numbered]{sn-jnl}

\usepackage{graphicx}%
\usepackage{multirow}%
\usepackage{amsmath,amssymb,amsfonts}%
\usepackage{amsthm}%
\usepackage{mathrsfs}%
\usepackage[title]{appendix}%
\usepackage{xcolor}%
\usepackage{textcomp}%
\usepackage{manyfoot}%
\usepackage{booktabs}%
\usepackage{algorithm}%
\usepackage{algorithmicx}%
\usepackage{algpseudocode}%
\usepackage{listings}%
\usepackage{braket}
\usepackage{tabularx}
\usepackage{booktabs}
\usepackage{caption}
\usepackage{siunitx}
\usepackage{tablefootnote}

\newcommand{\NNLOsat}{NNLO$_{\rm sat}$\;}
\newcommand{\NNLOgod}{$\Delta$NNLO$_{\rm GO}$(450)\;}
\newcommand{\NNLOgodlow}{$\Delta$NNLO$_{\rm GO}$(394)\;}

\begin{document}

\title[Low-energy electric dipole strength in $^8$He]{Low-energy dipole strength in $^8$He}

\author*[1]{\fnm{Francesca} \sur{Bonaiti}}\email{fbonaiti@uni-mainz.de}

\author[1, 2]{\fnm{Sonia} \sur{Bacca}}

\affil[1]{\orgdiv{Institut f\"ur Kernphysik and PRISMA$^+$ Cluster of Excellence}, \orgname{Johannes Gutenberg-Universit\"at}, \orgaddress{\postcode{55128}, \city{Mainz}, \country{Germany}}}

\affil[2]{\orgdiv{Helmholtz-Institut Mainz}, \orgname{Johannes Gutenberg-Universit\"at}, \orgaddress{\postcode{55128}, \city{Mainz}, \country{Germany}}}


\abstract{In this work, we present new ab initio coupled-cluster calculations of dipole-excited state properties of $^8$He based on the chiral effective field theory interaction 1.8/2.0 (EM). We focus on the dipole polarizability, and compare the results to our previous study [\href{https://journals.aps.org/prc/abstract/10.1103/PhysRevC.105.034313}{Phys. Rev. C 105, 034313 (2022)}] and subsequent theoretical work. With the aim of connecting the presence of low-lying dipole strength to structure properties of $^8$He, we compute the point-neutron radius, finding excellent agreement with available experimental data, and investigate its correlation with the dipole polarizability. }

\keywords{halo nuclei, dipole polarizability, electromagnetic observables, nuclear structure}

\maketitle

\section{Introduction}\label{sec1}

At the boundaries of the nuclear chart, the force binding protons and neutrons in the nucleus gives rise to intriguing phenomena, such as the emergence of halo structures, with weakly-bound nucleons orbiting a compact core at a distance. Halo nuclei are characterized by small separation energies and extended matter radii, which do not follow the typical $A^{1/3}$ scaling found in the valley of stability.

Among halo systems, $^8$He represents an interesting case study. It has the most extreme neutron-to-proton ratio ($N/Z = 3$) in the nuclear chart and up to now, it is the only known four-neutron halo. This unique configuration has been leveraged in a recent groundbreaking experiment, performed at RIKEN, where a $^{8}$He knockout reaction at large momentum transfer led to the observation of a correlated four-neutron state~\cite{duer2022}. While from the theory point of view the existence of a bound tetraneutron is disputed (see e.g. Ref.~\cite{marques2021} and references therein), new experiments have either been made \cite{faestermann2022} or they are being planned for \cite{duer_priv}. The debate around the interpretation of the RIKEN $4n$ signal has triggered new investigations of the neutron halo distribution and the possible correlations among the loosely-bound neutrons in $^8$He \cite{lazauskas2023,yamaguchi2023}.

In halo nuclei, the extended size of the neutron cloud is often mentioned together with a strong enhancement of the electric dipole response at low excitation energies. Revealed by early measurements with radioactive ion beams \cite{kobayashi1989}, this so-called "soft E1 excitation" ranks among the main features identifying halo nuclei \cite{aumann2013}. However, the presence of a soft dipole mode in $^8$He is still a controversial issue from both the experimental and theoretical point of view.

Coulomb excitation~\cite{meister2002} and nuclear fragmentation~\cite{markenroth2001} experiments first supported the existence of a $1^-$ resonance at an excitation energy of around $4$ MeV in the spectrum of $^8$He. Later, $(t,p)$ transfer reaction data~\cite{golovkov2009} confirmed the presence of low-lying dipole strength, but at a lower energy of $3$ MeV. In contrast, the Coulomb excitation experiment of Ref. \cite{iwata2000} attributed a relatively small fraction of the total energy-weighted dipole sum rule (less than 3$\%$) to a potential soft dipole mode. Also, a measurement of breakup of $^8$He on carbon \cite{xiao2012} found the spin-parity assignment of the excited state at $4$ MeV to $1^-$ highly uncertain. These results were validated by a recent inelastic proton scattering experiment \cite{holl2021}, where the measured angular distribution was found to be incompatible with a low-lying dipole resonance. Unpublished, high-statistics Coulomb excitation data obtained at RIKEN by the SAMURAI collaboration are expected to shed light on this controversy \cite{lehr2022}.

From the theory point of view, our recent work \cite{bonaiti2022} was the first to tackle dipole excited-state properties of $^8$He in the framework of $ab$ $initio$ calculations based on Chiral Effective Field Theory ($\chi$EFT) interactions \cite{epelbaum2009,machleidt2011,hammer2020}. Employing Coupled-Cluster (CC) theory \cite{hagen2014} merged with the Lorentz Integral Transform (LIT) technique \cite{efros2007} in the so-called LIT-CC approach \cite{bacca2013,bacca2014}, we observed low-energy dipole strength at around $5$ MeV. Subsequent works based on the cluster orbital shell model (COSM) \cite{myo2022,myo2022_ptep} and the equation-of-motion multiphonon approach (EMPM) \cite{degregorio2023} support the presence of a soft dipole mode.  In contrast, a random-phase approximation (RPA) calculation within a density functional theory (DFT) framework \cite{piekarewicz2022} disfavours such scenario, attributing it to spurious center-of-mass contaminations.

Motivated by this ongoing debate in both the theoretical and experimental communities, in this work we complement our previous analysis of $^8$He in two ways.  First, we provide new LIT-CC predictions of dipole excited-state properties of $^8$He employing the $\chi$EFT interaction 1.8/2.0 (EM) \cite{hebeler2011}. This interaction contains SRG-evolved two-nucleon forces at next to next to next to leading order (N3LO) and three-body forces at next to next to leading order (NNLO). The  nuclear force models used in Ref. \cite{bonaiti2022} (namely, \NNLOsat \cite{ekstrom2015}, \NNLOgod and \NNLOgodlow \cite{jiang2020}) are instead derived at NNLO in the chiral expansion. Second, using these four $\chi$EFT interactions, we investigate the halo structure of this nucleus by calculating the point-neutron radius and explore its correlation with the dipole polarizability.

The structure of this paper is the following. In Section 2, we give an overview of the LIT-CC method employed in this work. In Section 3, we present our new results, devoting one subsection to the discretized response function and dipole polarizability and one subsection to the analysis of correlations between dipole polarizability and point-neutron radius. In Section 4, we draw our conclusions.

\section{Review of the method}\label{sec2}
Electric dipole excitations of nuclei can be studied introducing the dipole response function, which is defined as:
\begin{equation}
   R(\omega) =  \sum_{\mu} |\braket{\Psi_{\mu}|\Theta|\Psi_0}|^2 \delta(E_{\mu} - E_0 - \omega) \,,
\label{response}
\end{equation}
where $\ket{\Psi_{\mu}}$ are the excited states obtained from the ground state $\ket{\Psi_0}$ applying the dipole operator $\Theta$, and $\omega$ is the photon energy.

To rewrite Eq. (\ref{response}) in the CC formalism  \cite{hagen2014}, we will briefly go through how $\ket{\Psi_0}$ and $\ket{\Psi_{\mu}}$ can be calculated in this approach. Given the nuclear Hamiltonian $H$, in CC theory one starts from a reference state $\ket{\Phi_0}$ expanded on the harmonic oscillator basis, and constructs the correlated many-body wavefunction via an exponential ansatz
\begin{equation}
    \ket{\Psi_0} = e^{T} \ket{\Phi_0}\,.
\label{exp_ansatz}
\end{equation}
The cluster operator $T$ can be written in terms of a sum of $n$-particle $n$-hole excitations, $T = T_1\;+\;T_2\;+\;T_3\;+ \dots +\;T_A$. Using Eq. (\ref{exp_ansatz}), the Schr\"odinger equation becomes
\begin{equation}
    H e^T  \ket{\Phi_0} = E_0 e^T  \ket{\Phi_0} \rightarrow \overline{H} \ket{\Phi_0} = E_0  \ket{\Phi_0}
\end{equation}
where $\overline{H} = e^{-T} H e^T$ is the similarity-transformed Hamiltonian. As the latter is non-Hermitian, left and right eigenstates of $\overline{H}$ are different. While the right eigenstate of $\overline{H}$ corresponds to the reference state $\ket{\Phi_0}$, the left eigenstate can be determined via the relation $\ket{\Psi_{0,L}} = (1 + \Lambda) \ket{\Phi_0}$, where $\Lambda$ is a de-excitation operator having a structure analogous to $T$.

Excited states are computed employing the equation-of-motion coupled-cluster (EOM-CC) method \cite{stanton1993}. The latter is based on the following ansatz for right and left excited states: 
\begin{equation}
    \ket{\Psi_{\mu}} = R_{\mu} e^T \ket{\Phi_0}, \quad \quad \bra{\Psi_{\mu}} =  \bra{\Phi_0} L_{\mu} e^{-T},  
    \label{eom-cc}
\end{equation}
where the EOM operators $R_\mu$ and $L_{\mu}$ can also be expanded in terms of particle-hole excitations. The EOM-CC ansatz allows to reformulate the computation of excited states as a right (left) matrix eigenvalue problem, with the excitation energies $E_{\mathrm{exc}, \mu} = E_{\mu} - E_0$ as eigenvalues and the amplitudes of $R_\mu$ ($L_{\mu}$) as eigenvectors. 

In CC theory, the particle-hole expansion of $T$ and $\Lambda$ for the ground state and of the EOM operators $R_{\mu}$ and $L_{\mu}$ for the excited states is truncated due to computational limitations. The most frequently employed approximation is coupled-cluster singles and doubles (CCSD), where we include up to 2p-2h excitations. Higher accuracy can be obtained within the CCSDT-1 framework, where leading order 3p-3h configurations are also present. In this work, we choose to adopt the same approximation scheme for the ground state and the excited state computations, either CCSD or CCSDT-1. 

Substituting Eq. (\ref{exp_ansatz}), (\ref{eom-cc}) in Eq. (\ref{response}), we can express $R(\omega)$ as
\begin{equation}
    R(\omega) = \sum_{\mu} \langle \Phi_0 \vert (1 + \Lambda ) \overline{\Theta}^\dagger R_\mu \vert \Phi_0 \rangle \langle \Phi_0 \vert L_\mu \overline{\Theta} \vert \Phi_0 \rangle \delta(E_{\mu} - E_0 - \omega)\,,
\end{equation}
where $\overline{\Theta} = e^{-T} \Theta e^T$ is the similarity-transformed dipole operator. 

We point out that the sum over $\mu$ in Eq. (\ref{response}) encompasses both bound and continuum excited states of the nucleus. This makes the calculation of response functions particularly challenging, as all the possible unbound configurations arising from the break-up of the nucleus at a given energy are involved. To avoid the issue of explicitly computing nuclear continuum wave functions, one can resort to the Lorentz Integral Transform (LIT) method. The LIT technique is based on the calculation of an integral transform with Lorentzian kernel of the response:
\begin{equation}
    L(\sigma,\Gamma) = \frac{\Gamma}{\pi} \int d\omega\; \frac{R(\omega)}{(\omega-\sigma)^2 + \Gamma^2}.
\label{transform}
\end{equation}
Within the so-called LIT-CC method \cite{bacca2013,bacca2014}, the LIT is connected to the solution of a EOM-CC equation with a source term. Once $L(\sigma,\Gamma)$ is determined, one can recover $R(\omega)$ via a numerical inversion procedure \cite{efros2007}.

Starting from the LIT, one can access the moments of the response function, i.e., the moments of $R(\omega)$ interpreted as a distribution function, which can be written as: 
\begin{equation}
    m_n = \int d\omega\;\omega^n R(\omega), 
\end{equation}
where $n$ is an integer. As the Lorentzian kernel tends to a Dirac delta function in the limit $\Gamma\rightarrow 0$, we get:
\begin{equation}
    L(\sigma, \Gamma\rightarrow 0) = \int d\omega\;R(\omega)\delta(\omega -\sigma) = R(\sigma).
\label{litsmallgamma}
\end{equation}
Taking advantage of Eq. (\ref{litsmallgamma}), we can directly compute the moments from the LIT as:
\begin{equation}
    m_n = \int d\sigma\; \sigma^n L(\sigma, \Gamma\rightarrow 0).
\label{mn_def}
\end{equation}
It is worth pointing out that this strategy does not require the inversion of $L(\sigma, \Gamma)$, which represents an additional source of uncertainty, and it has been proved to be equivalent to the integration of the response~\cite{miorelli2016}. 

In this work we will focus in particular on the inverse-energy-weighted dipole sum rule $m_{-1}$, which is proportional to the electric dipole polarizability $\alpha_D$: 
\begin{equation}
    \alpha_D = 2\alpha \int d\omega\; \frac{R(\omega)}{\omega} = 2\alpha m_{-1}.
    \label{alphaD}
\end{equation}

\section{Results}\label{sec3}
\subsection{Discretized response function and dipole polarizability}\label{sec31}
\begin{figure}[h!]
  \centering
  \includegraphics[scale=0.5]{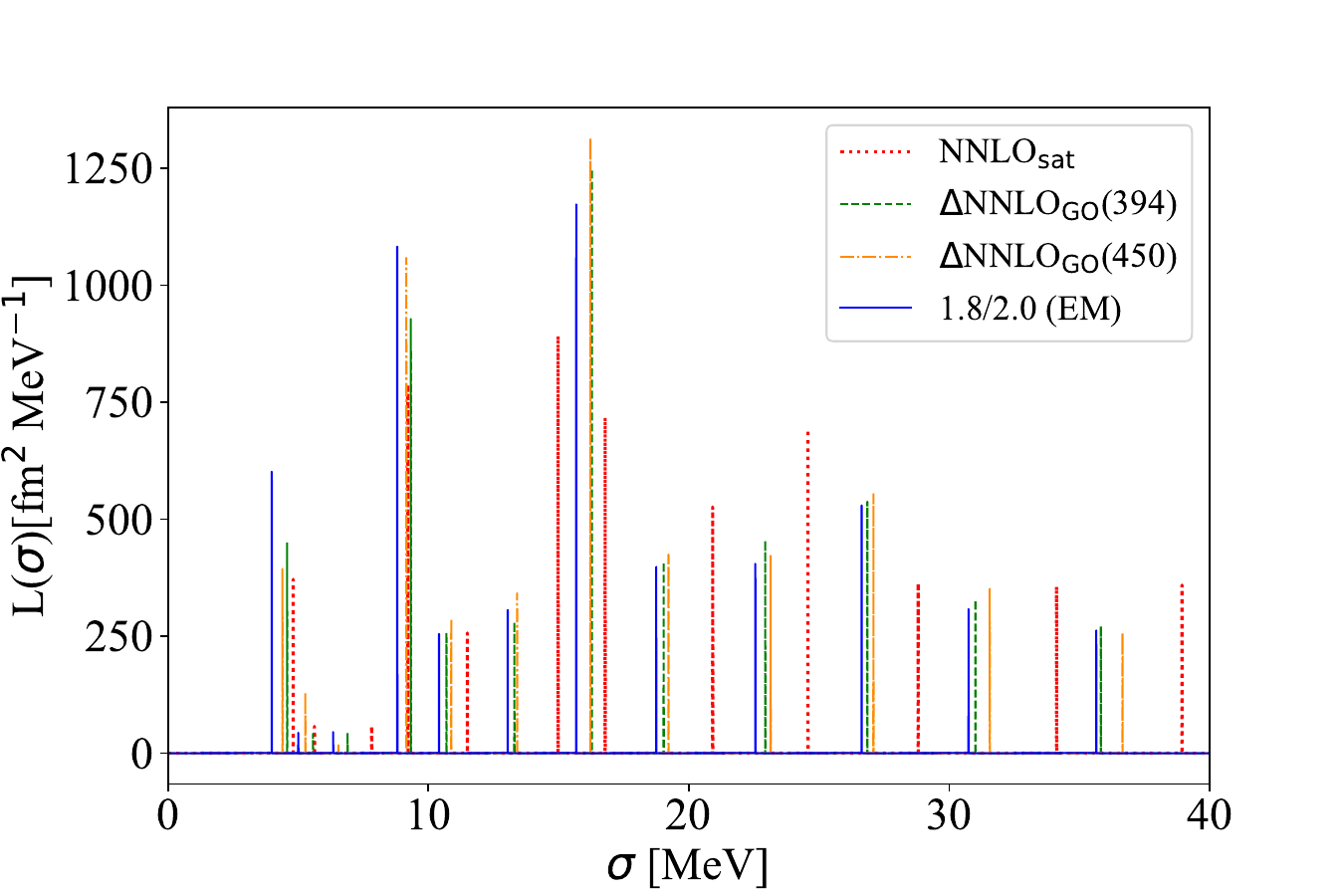}
  \caption{LIT of $^8$He with $\Gamma = 10^{-4}$ MeV in the CCSDT-1 framework for the chiral EFT interactions 1.8/2.0 (EM), $\Delta$NNLO$_{\mathrm{GO}}$(450), $\Delta$NNLO$_{\mathrm{GO}}$(394) and NNLO$_{\mathrm{sat}}$.}  
  \label{discretized_response}
\end{figure}

In this Section, we revisit our previous analysis of the dipole excited-state spectrum of $^8$He, by considering new calculations performed with the chiral force 1.8/2.0 (EM) \cite{hebeler2011}. We first consider the discretized dipole response function, that in our approach corresponds to the limit of the LIT for $\Gamma \rightarrow 0$. Fig. \ref{discretized_response} shows the LIT for $\Gamma = 10^{-4}$ MeV for the 1.8/2.0 (EM) interaction in comparison to the results obtained with the $\chi$EFT potentials used in our previous work \cite{bonaiti2022}. 

We observe that for the 1.8/2.0 (EM) interaction the response is characterized by low-energy strength emerging at slightly lower energies than predicted by the other $\chi$EFT interactions employed.
This reinforces the consistency of our coupled-cluster predictions with the experiments of Ref. \cite{meister2002,markenroth2001,golovkov2009}, as well as with the COSM \cite{myo2022,myo2022_ptep} and EMPM \cite{degregorio2023} theoretical calculations.

Starting from the discretized response function and employing Eq. (\ref{alphaD}), we can study the electric dipole polarizability $\alpha_D$. In Table \ref{table_alphaD}, the theoretical value of $\alpha_D$ for the 1.8/2.0 (EM) interaction is reported together with the results given by the set of chiral EFT Hamiltonians used in our previous work\footnote{The value quoted for \NNLOgodlow in our previous work was 0.39(2) fm$^3$, and it was obtained varying $\hbar\Omega$ between 12 and 16 MeV with the largest model space size available (N$_{\mathrm{max}}$ = 14). We update here our prediction on the basis of new calculations including $\hbar\Omega = 10$ MeV. We point out that the two values are compatible between errorbars. }. RPA \cite{piekarewicz2022} and EMPM \cite{degregorio2023} predictions of $\alpha_D$ are also included for comparison.
\begin{table}[ht]
\caption{Theoretical predictions for the dipole polarizability of $^8$He with the CCSDT-1 approximation for the four chiral EFT interactions under analysis. RPA predictions based on three relativistic mean-field (RMF) energy density functionals \cite{piekarewicz2022} and the EMPM result \cite{degregorio2023}, obtained with the \NNLOsat interaction, are also reported.}\label{table_alphaD}%
\begin{tabular}{l@{\hskip 1.5in}l}
\toprule
 Interaction   & $\alpha_D$ [fm$^3$] \\
\midrule
   1.8/2.0 (EM) & 0.48(3) \\
   \NNLOgod  & 0.42(3)  \\
   \NNLOgodlow  & 0.40(3)  \\
    \NNLOsat  & 0.37(3) \\\\
   RPA, RMF016 & 0.262\\
   RPA, RMF022 & 0.242\\
   RPA, RMF028 & 0.220\\
   EMPM, \NNLOsat  & 0.206 \\
\botrule

\end{tabular}
\end{table}

The uncertainty bands indicated in Table \ref{table_alphaD} for the coupled-cluster results take into account (i) the convergence with respect to the model space N$_{\mathrm{max}}$ and (ii) the effect of the many-body coupled cluster truncation. (i) is estimated considering the residual $\hbar\Omega$ dependence at the highest value of N$_{\mathrm{max}}$ available, corresponding to N$_{\mathrm{max}}$ = 14. (ii) is instead calculated comparing the two available CC approximation schemes, namely CCSD and CCSDT-1, and taking half of their difference as truncation uncertainty \cite{simonis2019}. (i) and (ii) are then summed in quadrature, leading to an uncertainty ranging between 6 and 8\% of the central value for the different interactions. In Ref. \cite{acharya2023} we discussed the uncertainty stemming from the truncation of the $\chi$EFT expansion on $\alpha_D$ for this nucleus. Focusing on the \NNLOgod model, NLO and NNLO results came out to agree with errorbars.

Looking at Table \ref{table_alphaD}, we notice that while $\alpha_D$ values from RPA \cite{piekarewicz2022} and EMPM \cite{degregorio2023} range between $0.20$ and $0.26$ fm$^3$, CC predictions are appreciably larger. Moreover, it is worth to point out the significant model dependence shown by our CC results for the polarizability of $^8$He. The \NNLOsat interaction, which performs well in reproducing the dipole polarizability in the medium-mass region \cite{fearick2023}, yields the lowest value for $\alpha_D$ in $^8$He. The 1.8/2.0 (EM) interaction delivers instead the largest prediction, which is more than $20\%$ higher than the \NNLOsat result. This is a consequence of the location of the low-lying dipole strength in $^8$He, which appears at slightly lower energies in the 1.8/2.0 (EM) case with respect to the other interactions, as shown in Fig. \ref{discretized_response}. Following the line of previous works as Ref. \cite{ekstrom2019}, a global sensitivity analysis of $\alpha_D$ in $^8$He would help in identifying which low-energy constants in the interaction are driving the substantial spread of the results obtained with different chiral forces. 

\subsection{Neutron radius and correlation with  dipole polarizability}\label{sec32}
To assess our understanding of the halo structure of $^8$He and its impact on low-energy dipole excitations, we calculate the point-neutron radius $R_n$ with the previously employed four $\chi$EFT interactions  and investigate correlations with the dipole polarizability.

\begin{figure}[h!]
 \centering
 \includegraphics[scale=0.5]{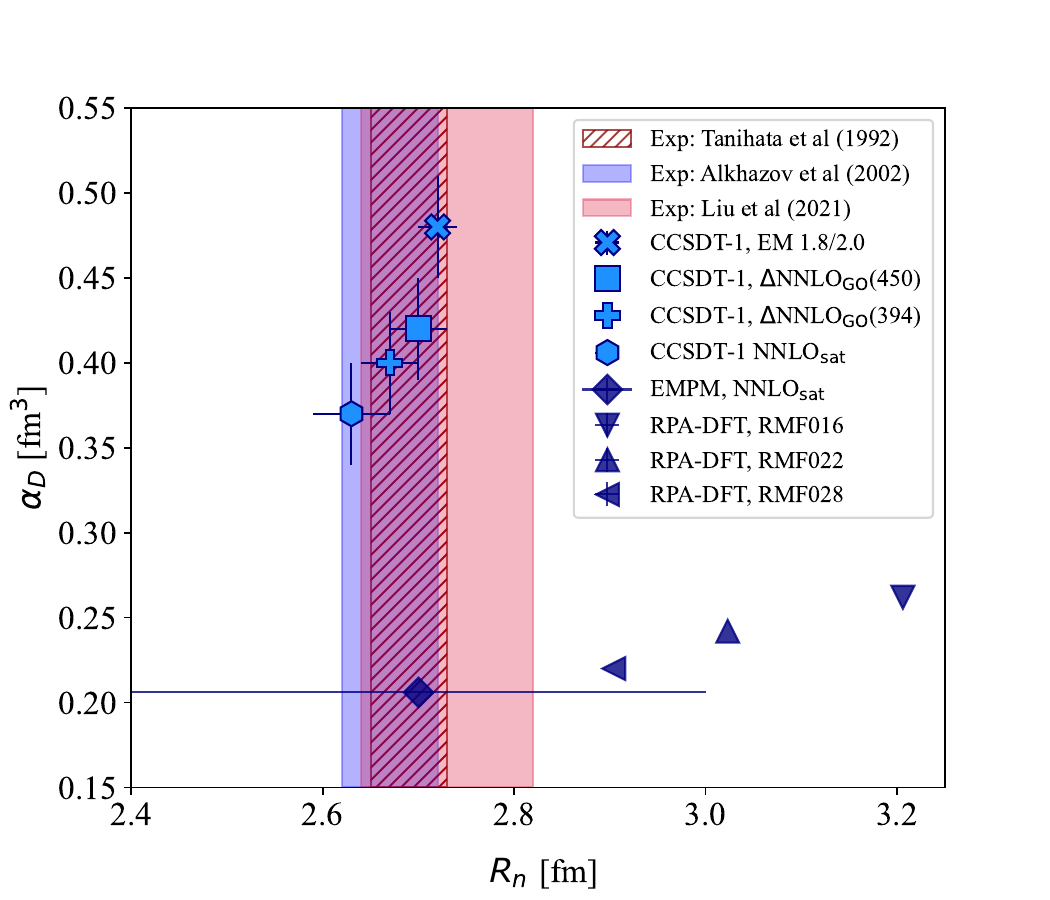}
 \caption{Electric dipole polarizability of $^8$He as a function of the corresponding point-neutron radius for the $\chi$EFT interactions 1.8/2.0 (EM), $\Delta$NNLO$_{\mathrm{GO}}$(450), $\Delta$NNLO$_{\mathrm{GO}}$(394) and NNLO$_{\mathrm{sat}}$. CC predictions for $\alpha_D$ and $R_n$ have been obtained in the CCSDT-1 approximation scheme. Comparison with the experimental determinations of Ref. \cite{tanihata1992,alkhazov2002,liu2021} as well as with the RPA-DFT \cite{piekarewicz2022} and EMPM \cite{degregorio2023} results is shown.}  
  \label{alphaD_rn}
\end{figure}

Fig. \ref{alphaD_rn} shows coupled-cluster predictions for $\alpha_D$ as a function of $R_n$, in comparison to three experimental determinations of $R_n$ \cite{tanihata1992,alkhazov2002,liu2021} and the theoretical estimates obtained by the RPA-DFT \cite{piekarewicz2022} and EMPM \cite{degregorio2023} approaches.
Experimental determinations of $R_n$ have been obtained on the basis of interaction cross section measurements in inverse kinematics on carbon \cite{tanihata1992} and hydrogen targets \cite{alkhazov2002,liu2021}, respectively. A Glauber analysis of the experimental data allows for the extraction of the nuclear matter radius $R_m$, which in combination with the point-proton radius $R_p$ can be employed to estimate $R_n$ according to the relation $R_m^2 = (Z R_p^2 + N R_n^2)/A$.

We find that our theoretical predictions for $R_n$ are in excellent agreement with all three experimental determinations. We estimated our theoretical uncertainty following the procedure outlined in Sect. \ref{sec31}. The residual dependence on convergence parameters and the truncation of the many-body CC expansion lead to a combined errorband varying between $1$ and $2\%$ for the different interactions. The EMPM value for $R_n$ is also in good accordance with experiment, while DFT results overestimate the data.

Looking at our predictions for different chiral forces, we notice the presence of a correlation between $\alpha_D$ and $R_n$, which emerges also from the RPA-DFT calculations of Ref. \cite{piekarewicz2022}. This may reflect a link between the low-lying dipole strength, which mainly determines $\alpha_D$, and an excitation of the weakly-bound excess neutrons. We can arrive to similar conclusions also for neutron-rich nuclei in the medium-mass region, as $^{48}$Ca \cite{simonis2019} and $^{68}$Ni \cite{kaufmann2020}. In fact, Refs. \cite{simonis2019,kaufmann2020} highlight the presence of a correlation between $\alpha_D$ and the point-proton radius $R_p$ for these two nuclei. At the same time, the authors also observed that increasing values of $R_p$ were associated to a corresponding rise in $R_n$, leading as a consequence to the emergence of a correlation between $\alpha_D$ and $R_n$.

\section{Conclusions}\label{sec4}
Motivated by recent theoretical and experimental interest towards the possible presence of a soft dipole mode in $^8$He, we reconsider the analysis of our previous work \cite{bonaiti2022} in the light of new coupled-cluster calculations of the discretized response function and dipole polarizability with the chiral force 1.8/2.0 (EM). As a difference with respect to the Hamiltonian models employed in Ref. \cite{bonaiti2022}, derived at NNLO in the chiral expansion, 1.8/2.0 (EM) contains also the SRG-evolved two-nucleon forces at N3LO.

The 1.8/2.0 (EM) interaction confirms the presence of low-lying dipole strength in $^8$He at around 5 MeV, and predicts the largest value for the polarizability, corresponding to $0.48(3)$ fm$^3$. The presence of low-energy dipole strength in $^8$He still remains an open problem which calls for further theoretical and experimental investigation. The new Coulomb excitation data from RIKEN \cite{lehr2022} are expected to sort out the soft dipole mode controversy and serve as a benchmark for theoretical predictions of $\alpha_D$.

Moreover, employing the 1.8/2.0 (EM) interaction and the set of chiral forces from our previous study, we computed the point-neutron radius, obtaining excellent agreement with experiment. We pointed out the presence of a correlation between dipole polarizability and point-neutron radius, which indicates the sensitivity of the low-energy dipole strength to the extent of the pont-neutron distribution in such an exotic halo nucleus. 

\backmatter

\bmhead{Acknowledgments} We would like to thank Gaute Hagen for access to the spherical coupled-cluster code. F.B. would also like to acknowledge interesting discussions with Tobias Frederico, Chlo\"e Hebborn and Ritu Kanungo. This work was supported by the Deutsche Forschungsgemeinschaft  (DFG, German Research Foundation) through Project-ID 279384907 - SFB 1245 and through the Cluster of Excellence "Precision Physics, Fundamental Interactions, and Structure of Matter" (PRISMA$^+$ EXC 2118/1, Project ID 39083149). This research used resources of the Oak Ridge Leadership Computing Facility located at ORNL, which is supported by the Office of Science of the Department of Energy under Contract No. DE-AC05-00OR22725, and of the supercomputer Mogon at Johannes Gutenberg Universit\"at Mainz.

\bibliography{sn-bibliography}

\end{document}